\documentclass[letter,twocolumn]{jpsj3}
\usepackage{graphicx}

\newcommand{\simj}{\stackrel{>}{_\sim}}

\begin{document}
\title{Superconductivity of Carbon Compounds with Sodalite Structure }

\author{Kazuhiro Sano, Yoshimi Masuda,  and Hiroki Ito}
\inst{Department of Physics Engineering, Mie University, Tsu, Mie 514-8507, Japan}

\abst{
We  investigate  the superconductivity of carbon compounds with  a sodalite structure, which  are similar to  hydrogen compounds  showing the high-temperature superconductivity.
A systematic analysis by  first-principles calculations is carried out, including examination of mechanical and dynamic instabilities under external pressure $P$.
These instabilities are classified on the phase diagram for   the effective doping charge  versus the lattice constant of the system. 
We also present  the superconducting  transition temperature $T_{\rm c}$ as a function of  $P$ for many carbon compounds and a pure  carbon system with the sodalite structure. 
Some of them  have  $T_{\rm c}$ of up to  about  100 K  at $P \simj 30$  GPa, and  the results suggest that the sodalite structure of carbon may be a key to producing phonon-mediated high-$T_{\rm c}$  superconductivity.
}

\maketitle
Since the high-temperature superconductivity (HTS) of metallic   hydrogen  was predicted at an extremely high pressure,\cite{Ashcroft-1968}  much effort has been made to clarify  its superconductivity  and/or to find new  hydrogen compounds  relevant to  metallic  hydrogen.\cite{McMahon2012,Y-Li2014,Drozdov-LaH10}
In particular, hydrogen compounds with a sodalite structure, such as  YH$_6$  are  predicted to have a transition temperature $T_{\rm c}$ of over 250 K\cite{Wang-2012, Y-Li-2015,  Heil-2019}.
On the basis of these predictions, experiments\cite{Troyan-2019-YH6,Kong-2019-YH6} were conducted. It was verified that the  $T_{\rm c}$s of these superconductors  are  close to those obtained by  first-principles calculations.

In  hydrogen compound superconductors,  high pressure may stabilize their characteristic structures and leads to a high phonon frequency.
In experiments, it is not easy  to generate  high pressure of over a few  hundred  GPa,  and   HTS of hydrogen compounds can only be achieved in  diamond anvil cells. 
It is desirable to reduce the required  pressure to realize HTS.

HTS of hydrogen compounds is  caused by the  phonon mediated attraction and its mechanism of the superconductivity is conventional. 
In this case,  $T_{\rm c}$ is governed by mainly two parameters, namely,  the characteristic phonon  frequency of the system $\omega_{\rm log}$ and  the  electron-phonon coupling constant $\lambda$.  When both parameters are large, HTS can be expected. 
In fact,  many hydrogen compounds with a sodalite structure have  a large  $\lambda$, which is larger than about 2.0.\cite{Wang-2012, Y-Li-2015, Heil-2019}

If the sodalite structure plays an important role in producing a large $\lambda$, it might be interesting to replace   the hydrogen atom of the sodalite structure  with another  atom.
A material formed by carbon atoms such as diamond is stable  at atmospheric pressure and   has a high phonon frequency  up to 2000 K.
%
%
\begin{figure}[hb]
\begin{center}
\includegraphics[width=0.5 \linewidth]{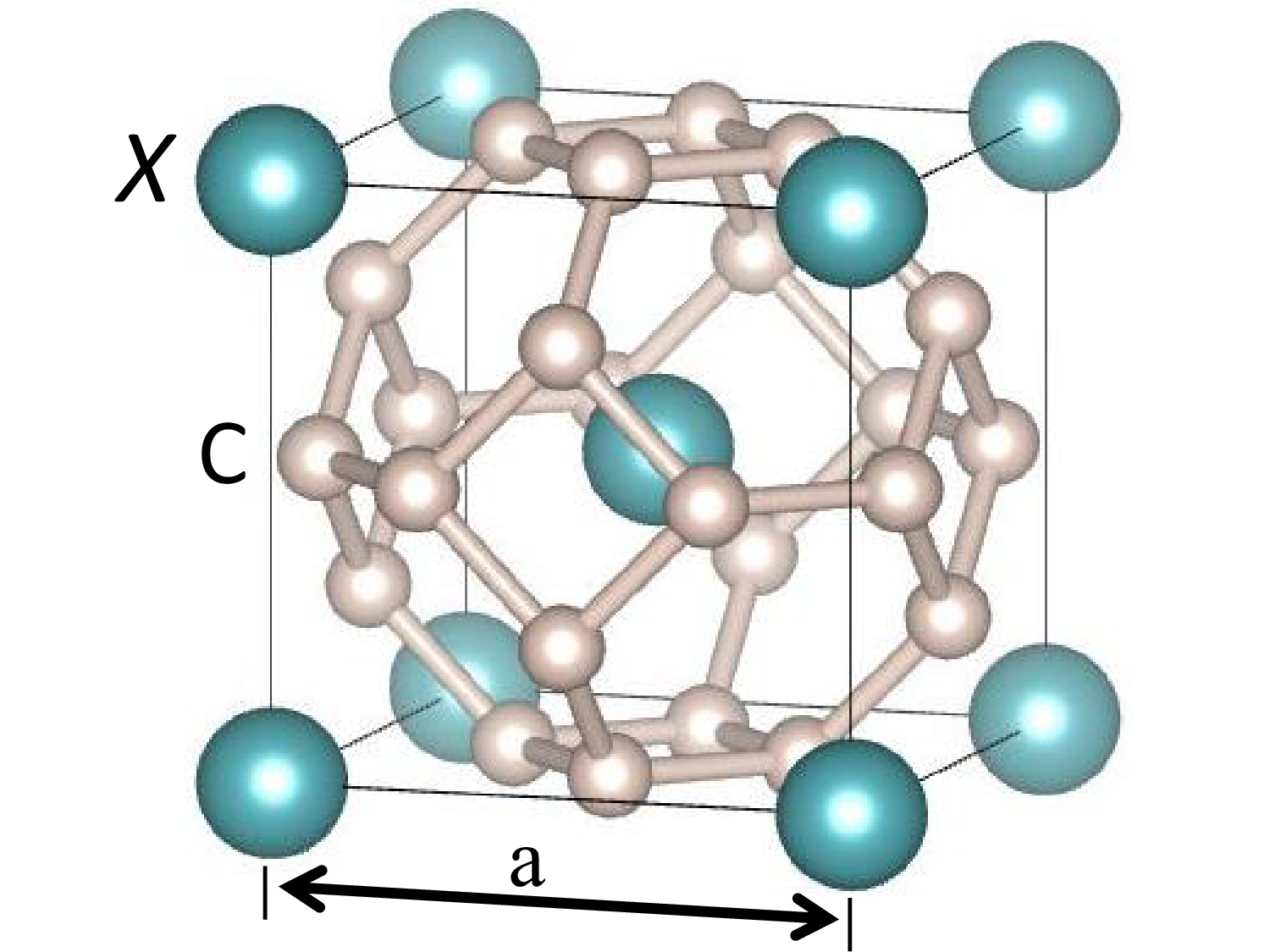}
\end{center}
\caption{(Color online)
   Structure of  the sodalite-type compound $X$C$_6$, where large spheres represent $X$ atoms and small spheres are carbon atoms.  Here, $a$ is the length of one side of the unit cube  consisting of two unit cells.
}
\label{model}
\end{figure}
Therefore, one way to reduce the required pressure and to realize HTS is  to replace  hydrogen   with  carbon.
In fact,  boron-doped diamond  has been  studied  as  a candidate material with   phonon-mediated HTS at atmospheric pressure.\cite{Kawano,Ma}

A sodalite structure with only carbon atoms, which is an insulator with a large charge gap, has already been predicted on the basis of the density functional theory.\cite{Pokropivny}
By combining it with other elements, we expect high $\omega_{\rm log}$ and large $\lambda $  as well as a hydrogen compound superconductor. 
In fact, it has been studied as a compound consisting of six carbon atoms and another atom, which is represented by $X$C$ _6$, where $X$ is Li, Na, Cl, and so on.\cite{Lu-2016,Khan2022}
Despite being under atmospheric pressure,  $T_{\rm c}$ is claimed to be up to 100 K, where  $\omega_{\rm log} \simeq 400$ K and   $\lambda \simeq 2.8$  for NaC$_6$.\cite{Lu-2016}
However, the existence of $X$C$ _6$ has  not yet been experimentally demonstrated  and its nature has   not been clarified. 
In addition, recent analysis of mechanical stability shows that some $X$C$_6 $ systems are unstable  at atmospheric pressure.\cite{Khan2022,Wei-2016}

 In this work,  we investigate  the superconductivity of  carbon compounds with  sodalite structure  using the first-principles calculations.
 Both of   mechanical and  dynamical instabilities are also examined by varying the pressure systematically.
Using the McMillan formulation\cite{McMillan,Allen},  we show  $T_{\rm c}$s of a pure C$ _6$ system  with fictitious charge and  many compounds as a function of  external pressure $P$.
These results would clarify  the stability and superconducting properties of $X$C$ _6$ systems. 


 In Fig. \ref{model}, we show the   structure of   $X$C$_6$, where $X$ stands for an atom bonded to  carbon atoms that form the skeleton of  the sodalite  structure.
Calculations are performed using  'Quantum ESPRESSO'(QE), which is an integrated software program of Open-Source computer codes for electronic-structure calculations.\cite{QE} 
In our calculation, we mainly use a $24 \times 24 \times 24$  Monkhorst-Pack grid for the electronic Brillouin zone integration and  a $4 \times 4 \times 4$ mesh  for phonon calculation.     
The elastic constants $c_{ij}$  are calculated by using '$\rm{Thermo\_pw}$'  which is a driver of QE routines\cite{Thermo}.  

 %
\begin{figure}[hb]
\begin{center}
\includegraphics[width=0.9 \linewidth]{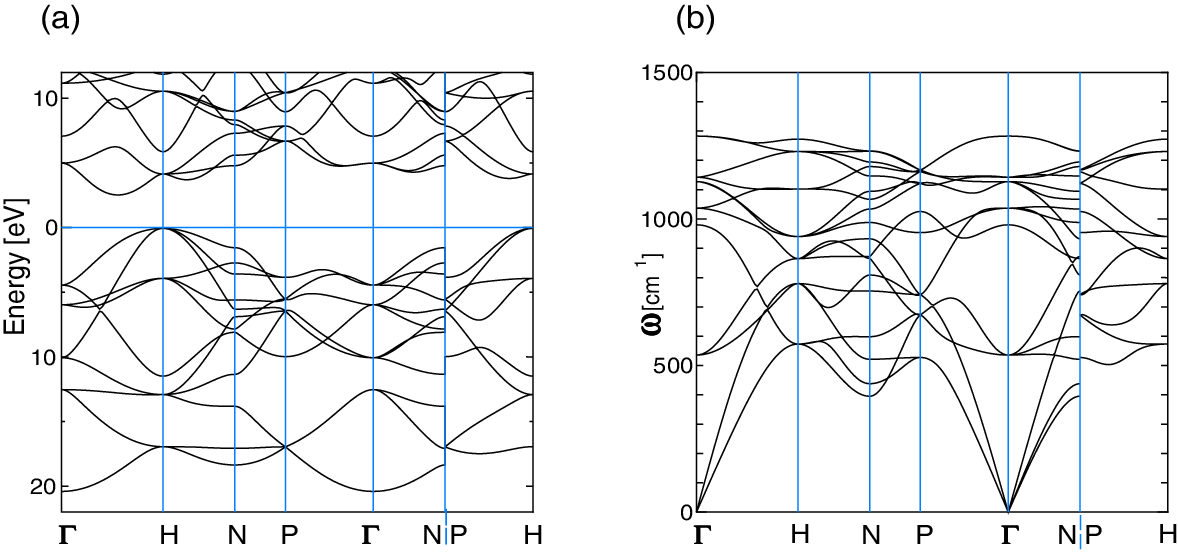}
\end{center}
\caption{(Color online)
(a) Band structure of  C$_6$ at zero pressure, where the  Fermi energy is  set to be zero.
(b)  Phonon dispersion of  C$_6$. 
 }
\label{C6-band}
\end{figure}
%
%
\begin{figure}[hb]
\begin{center}
\includegraphics[width=0.7 \linewidth]{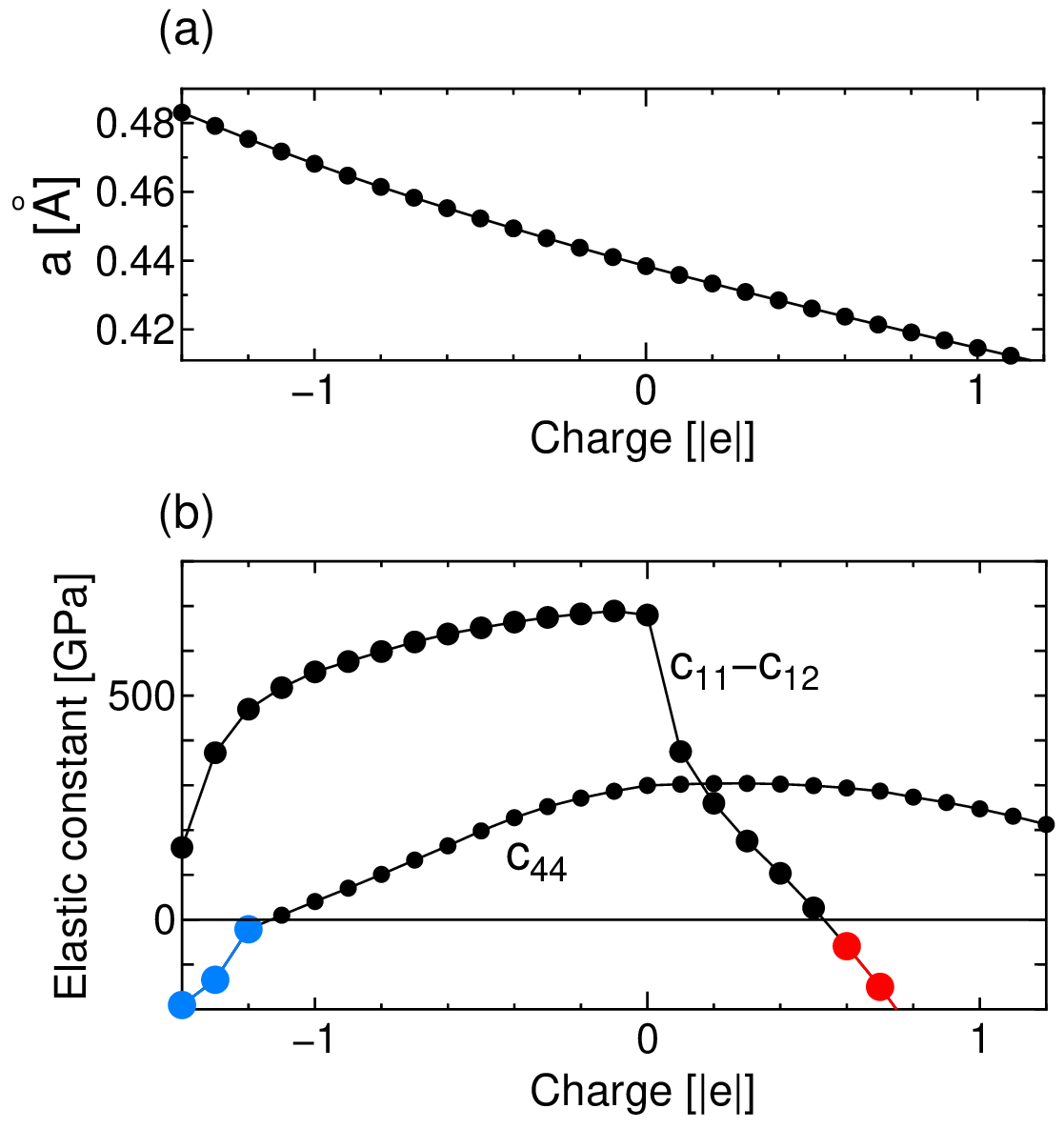}
\end{center}
\caption{(Color online)
(a) Lattice constant  $a$ of C$_6$ as a function of the fictitious charge. 
(b) The difference of elastic constants $c_{11}-c_{12}$ and  $c_{44}$    as  functions of the fictitious charge.
Here, the large blue solid circles represent mechanical instability points corresponding to  $c_{44} <0$, and the large red solid circles represent mechanical instability points corresponding to $c_{11}-c_{12} <0$.
}
\label{C6elastic}
\end{figure}
%
%
\begin{figure}[th]
\begin{center}
\includegraphics[width=1.0 \linewidth]{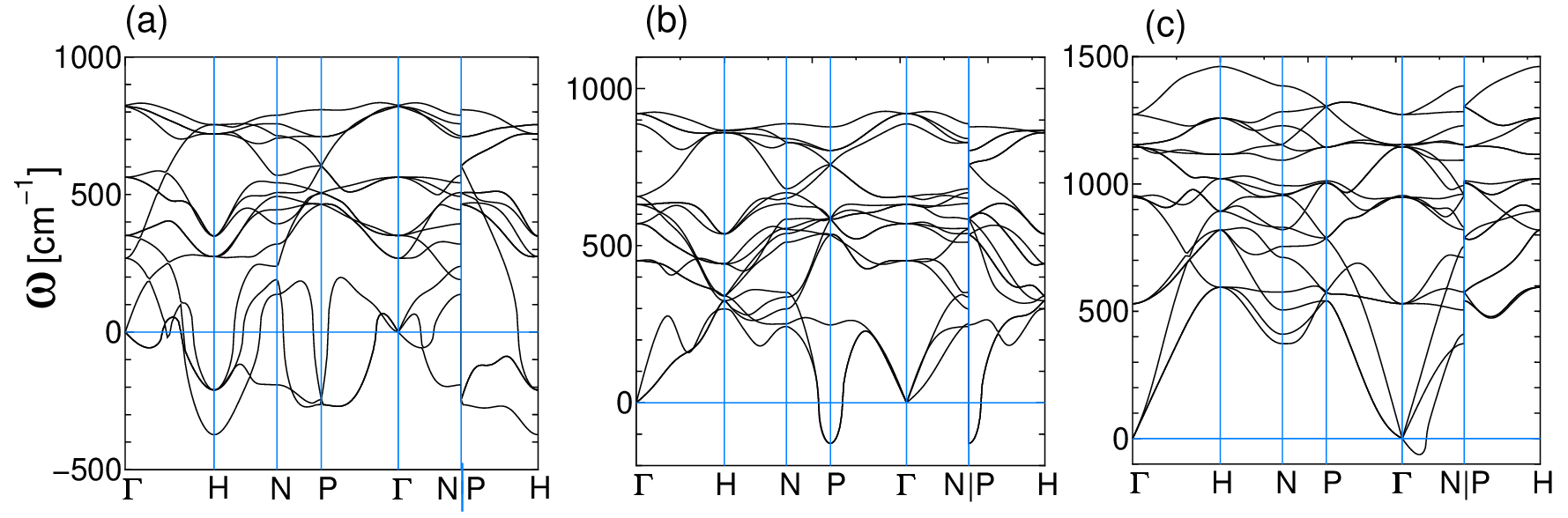}
\end{center}
\caption{(Color online)
(a) Phonon dispersion of  C$_6$,  where the  fictitious charge  is  set to  -1.4 $|e|$.
(b)   Phonon dispersion of  C$_6$,  where the  fictitious charge  is  set to  -1.0 $|e|$.
(c)   Phonon dispersion of  C$_6$,  where the  fictitious charge  is  set to  0.6 $|e|$.
}
\label{C6dope-ph}
\end{figure}
%
First, we consider a pure carbon  system with the sodalite structure, C$_6$. It  forms the skeleton of this superconductor and may be the basis for  understanding the superconductivity of  $X$C$_6$.
In Fig. \ref{C6-band}(a), we show the band structure of an electron at $P=0$ GPa.
It indicates that C$_6$ is an insulator with a charge gap of about  2.5 eV, which agrees with the result of a previous work.\cite{Pokropivny} 

Figure \ref{C6-band}(b) shows that   the phonon dispersion  becomes positive at any point in the Brillouin zone and the system is stable against  phonon excitation at $P=0$ GPa.
It indicates that the system becomes dynamically stable.
We  also confirmed that the system can withstand a pressure of at least 250 GPa.
%
%
%
To examine the relationship between  the instability and  electronic state of  C$_6$,  we apply  the rigid-band approximation\cite{Subedi,Sano-2019} to  C$_6$.
In this method,  a fictitious charge is introduced   into the target system by assuming a rigid band. It changes only the Fermi energy of the system according to carrier density.  
 It will be useful for  analyzing the electronic state  of  $X$C$_6$,  because the compound $X$C$_6$  may be considered as a system with an $X$-atom doped  into  C$_6$.
For example,   if $X$ is an alkali metal, it would  correspond to the case of  electron doping and  the Fermi energy  of the system is increased. 

 In Fig. \ref{C6elastic}(a), we show the lattice constant  $a$ of C$_6$ as a function of the fictitious charge, whose value  is normalized  by the absolute charge value of  an electron,  $|e|$. 
The figure indicates that the electron doping  increases  the lattice constant $a$  of the unit  cube.
In other words, the effect of carrier doping may be equivalent to applying   pressure to the system as a side effect.
 In Fig. \ref{C6elastic}(b),  the difference of elastic constants $c_{11}-c_{12}$ and  $c_{44}$ as  functions of the fictitious charge are shown.
It is known that  conditions   $c_{11}-c_{12} < 0$ and  $c_{44} <0$  lead  to mechanical instability of the system.\cite{Born,Grimvall,Mouhat}
The former means a negative Young's modulus and the latter corresponds to negative rigidity.

Here, we consider the relationship between  mechanical and  dynamical instabilities.
Figures  \ref{C6dope-ph}(a)-(c)  show the phonon dispersion of C$_6$,  where  the fictitious charges  are set to   -1.4 $|e|$,   -1.0 $|e|$  and   0.6 $|e|$, respectively.
In the cases of Figs. \ref{C6dope-ph}(a) and \ref{C6dope-ph}(c),  the dynamical instability occurs as well as  the mechanical  instability, as shown Fig. \ref{C6elastic}(b).
 From the  view point of instability, both results of   the dynamical and  mechanical  instabilities seem to be consistent with each other. 
However, Fig. \ref{C6dope-ph}(b) shows that a negative frequency occurs at around the P-point.
Therefore, when the fictitious charge  is -1.0 $|e|$,  dynamical instability occurs although  mechanical instability does not appear. 
This suggests that the analysis of  dynamic instability provides more information  than that of  mechanical instability.
Since the long-wavelength limit of lattice vibrations corresponds to  uniform deformation of a  whole system,  phonon dispersion with a negative frequency   at around the $\Gamma$-point  may lead  to   mechanical instability. 
However, the analysis of  mechanical instability is not  applicable to the analysis of  instability of deformation characterized by finite wavelength such as that at the P-point. 
Therefore, the scope of   dynamic instability is beyond that of  mechanical instability.\cite{Grimvall}
%
%
\begin{figure}[th]
\begin{center}
\includegraphics[width=0.6 \linewidth]{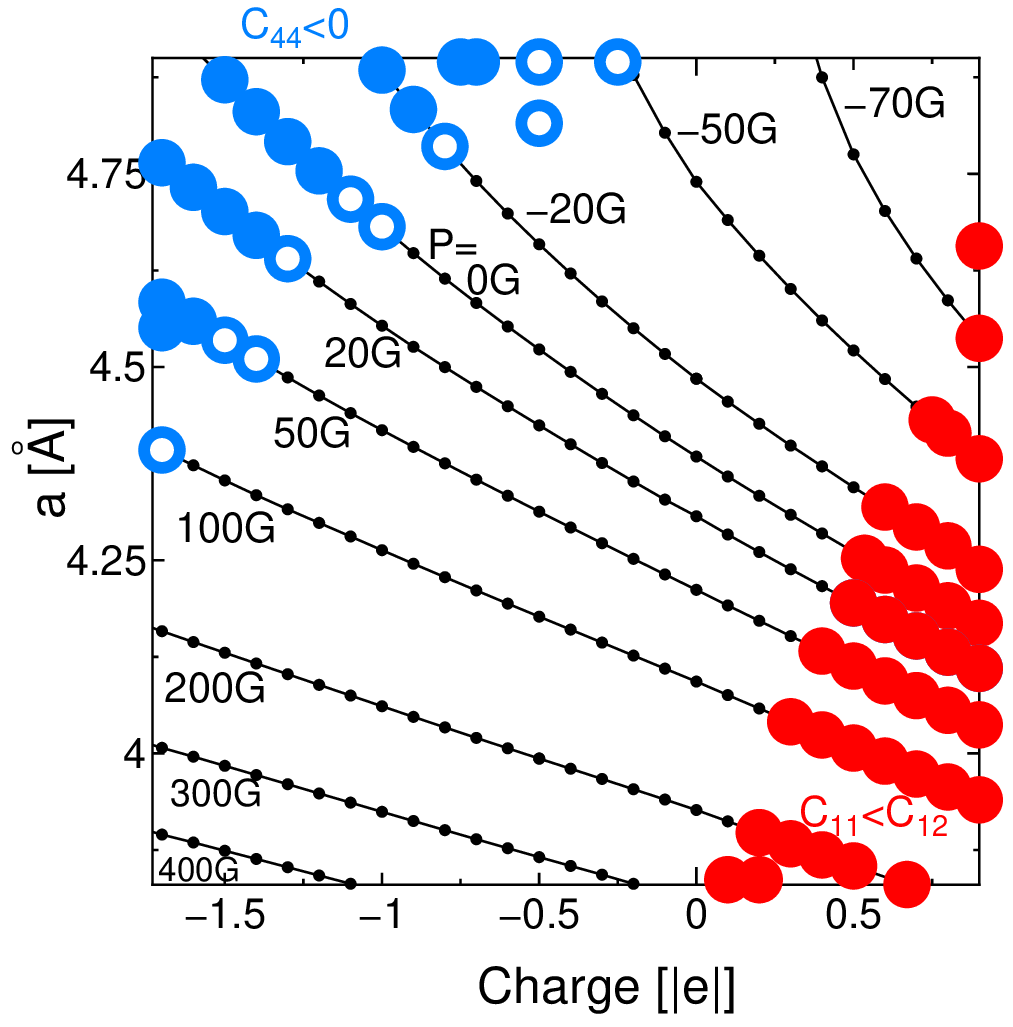}
\end{center}
\caption{(Color online) Phase diagram  on  a plane  of  the fictitious charge versus the lattice constant $a$,  where each small point is classified by pressure $P$.
Large blue solid circles represent mechanical instability points corresponding to the region of  ${\rm c}_{44} < 0$, and  large blue open circles   represent not  mechanical but dynamical instability points.
Furthermore, large red solid circles represent mechanical instability points corresponding to the region of ${\rm c}_{11} < {\rm c}_{12}$.
  }
\label{phase-dia-A}
\end{figure}
%
\begin{figure}[ht]
\begin{center}
\includegraphics[width=1.0 \linewidth]{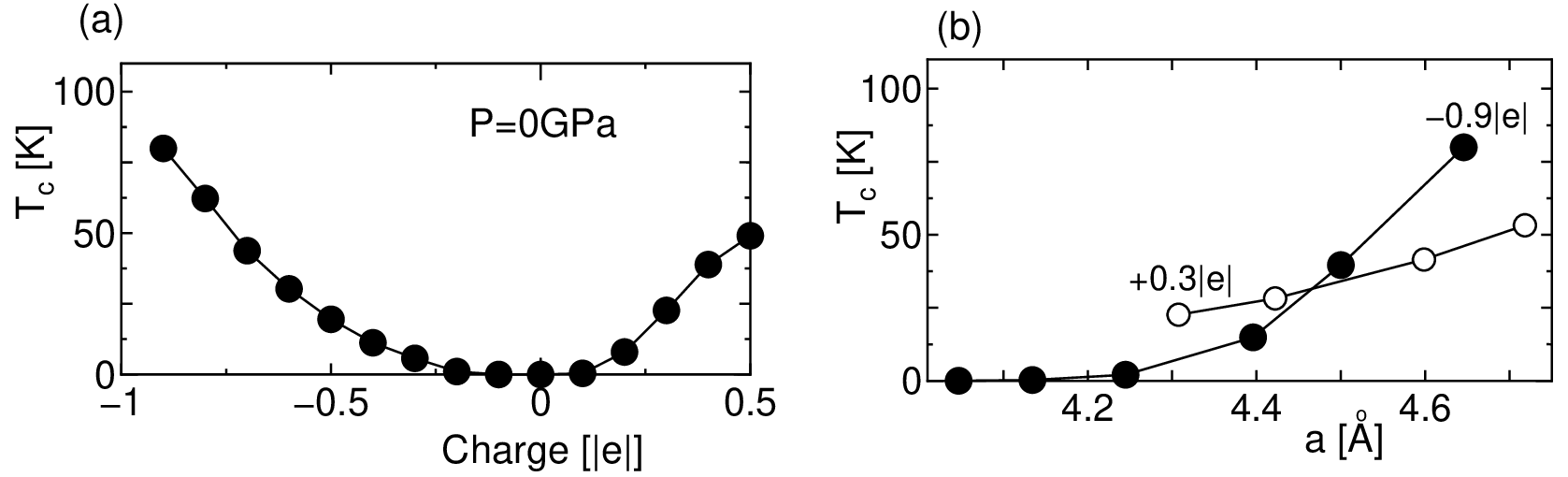}
\end{center}
\caption{
 (a) $T_{\rm c}$ of superconductivity of C$_6$ as a function of fictitious charge at $P=0$ GPa.
(b) $T_{\rm c}$ of  C$_6$ as a function of the lattice constant $a$ for the systems with fictitious charges -0.9$|e|$ (solid circles) and +0.3$|e|$ (open circles).
}
\label{C6super}
\end{figure}

In Fig. \ref{phase-dia-A}, we show a phase diagram of  C$_6$. Here, the horizontal axis is  the fictitious charge and  the vertical axis  is the lattice constant $a$.
In this figure, the lower right area  with large red solid circles   shows  the mechanical instability  characterized by the inequality $c_{11} - c_{12} < 0$. On the other hand,  the upper left area with large blue solid circles shows  the mechanical instability characterized by the inequality $c_{44} < 0$.
 In addition, large blue open circles   represent the area indicating  only dynamical instability, not mechanical instability.

Although the doped C$_6$ system is a fictitious model, it is interesting   to calculate $T_{\rm c}$.\cite{Lu-2016} 
 The result may become a good reference for considering the superconductivity of $X$C$_6$ systems. 
We show $T_{\rm c}$\cite{Tc-too-high} as a function of the fictitious charge and  the lattice constant  $a$ for C$_6$ in  Fig. \ref{C6super}.
Figure \ref{C6super}(a) indicates that $T_{\rm c}$ increases with the absolute value of the fictitious charge.
 It also shows that $T_{\rm c}$ reaches  $\sim 80$ K at -0.9 $|e|$,  where $\lambda$  is  1.94 and  $\omega_{\rm log}$ is 471  K. 
Figure \ref{C6super}(b) shows that $T_{\rm c}$ increases with  $a$, and this tendency seems to be unrelated to  the sign  of the fictitious charge.
%
\begin{figure}[th]
\begin{center}
\includegraphics[width=1.0 \linewidth]{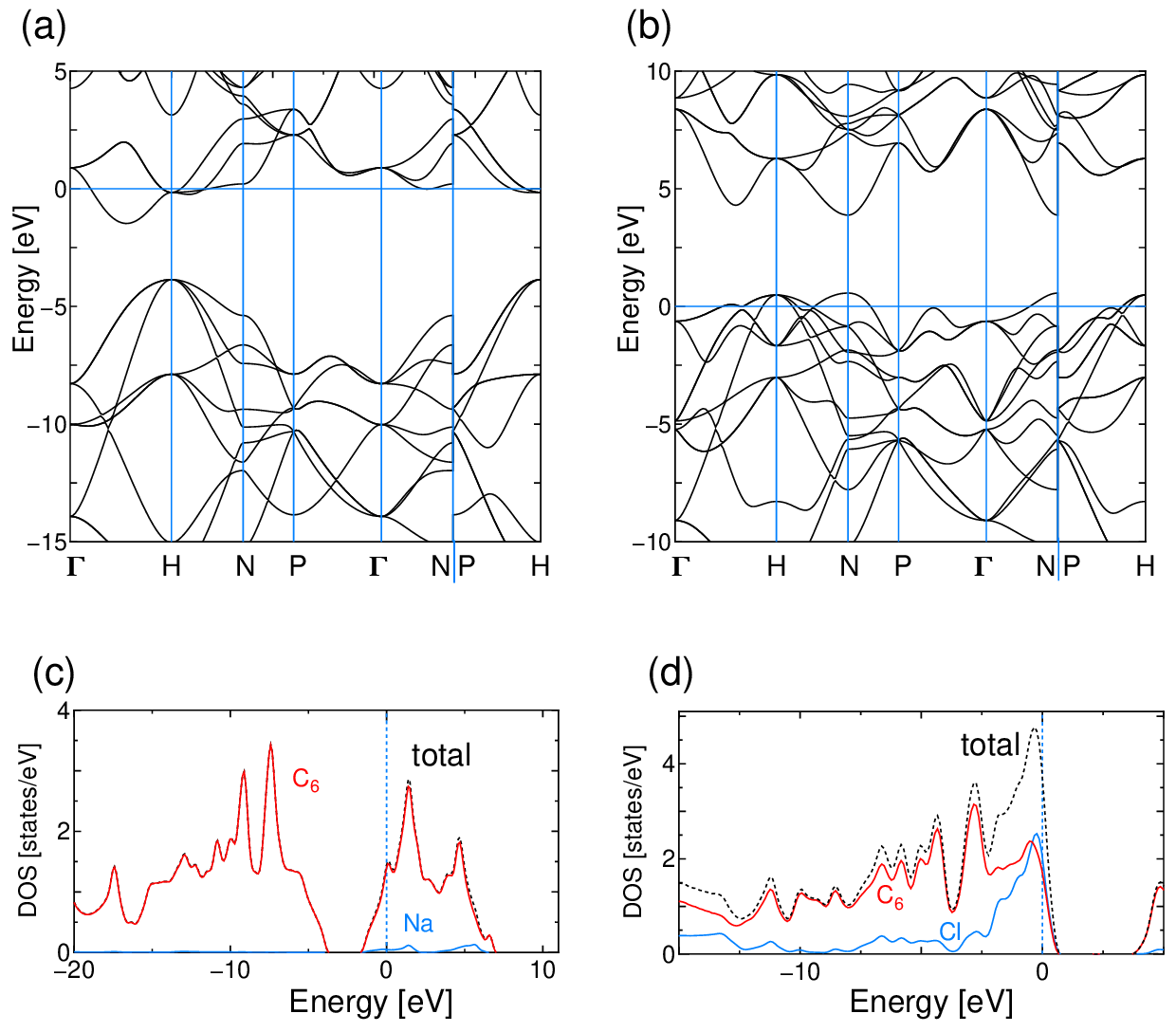}
\end{center}
\caption{(Color online) Band structures of (a)  NaC$_6$ at 30 GPa and (b) ClC$_6$ at 0 GPa, where the Fermi energy is set to be zero.
DOSs of electrons  for (c)  NaC$_6$ at 30 GPa and (d) ClC$_6$ at 0 GPa, where the components decomposed for each atom are color-coded.
 }
\label{e-band-all3}
\end{figure}
%
\begin{figure}[th]
\begin{center}
\includegraphics[width=1.0 \linewidth]{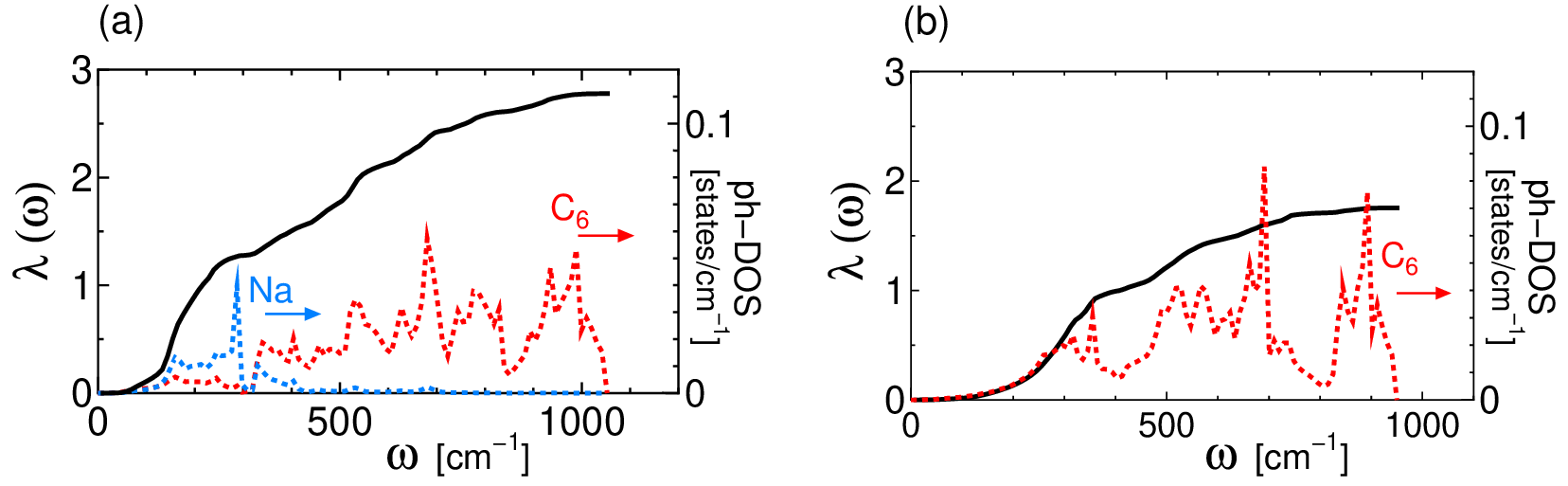}
\end{center}

\caption{(Color online)  Electron-phonon coupling $\lambda (\omega)$ as a function of $\omega$  for (a)  NaC$_6$ at 30 GPa and (b) C$_6$ with the fictitious charge of -0.9 $|e|$  at 0 GPa.
Broken lines indicate phonon(ph)-DOS, where the components of ph-DOS decomposed for each atom are color-coded.
 }
\label{a2F}
\end{figure}

\begin{figure}[th]
\begin{center}
\includegraphics[width=1.1 \linewidth]{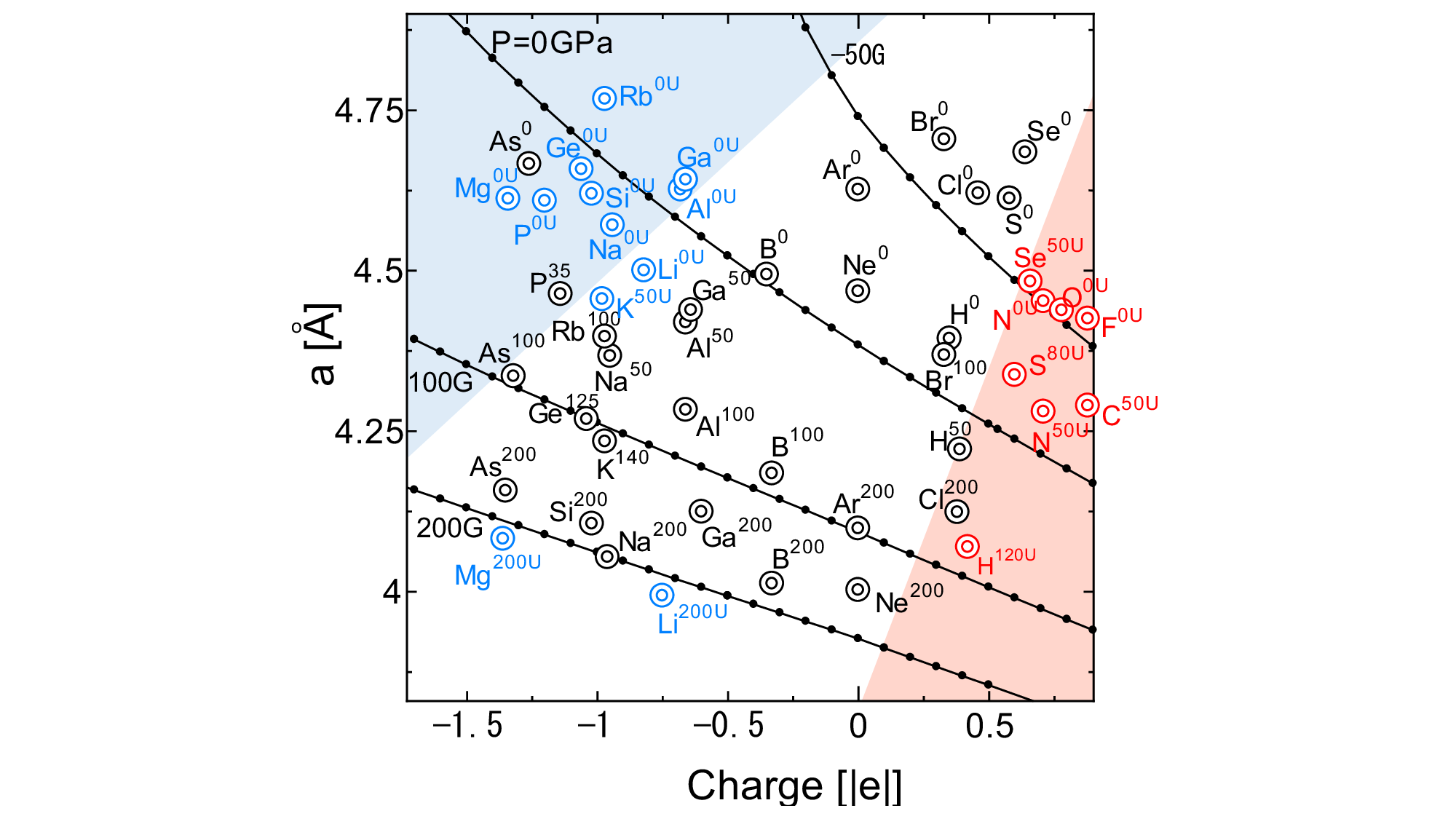}
\end{center}
\caption{(Color online)  Corresponding points of $X$C$_6$ on the phase diagram of  Fig. \ref{phase-dia-A},  which are  represented by double circles.
The superscript of each double circle indicates the pressure of $X$C$_6$ in the unit of  GPa, and the additional superscript `U`   means  its mechanical and/or dynamical instability.
}
\label{phase-dia-B}
\end{figure}
%
%
\begin{figure}[ht]
\begin{center}
\includegraphics[width=0.7 \linewidth]{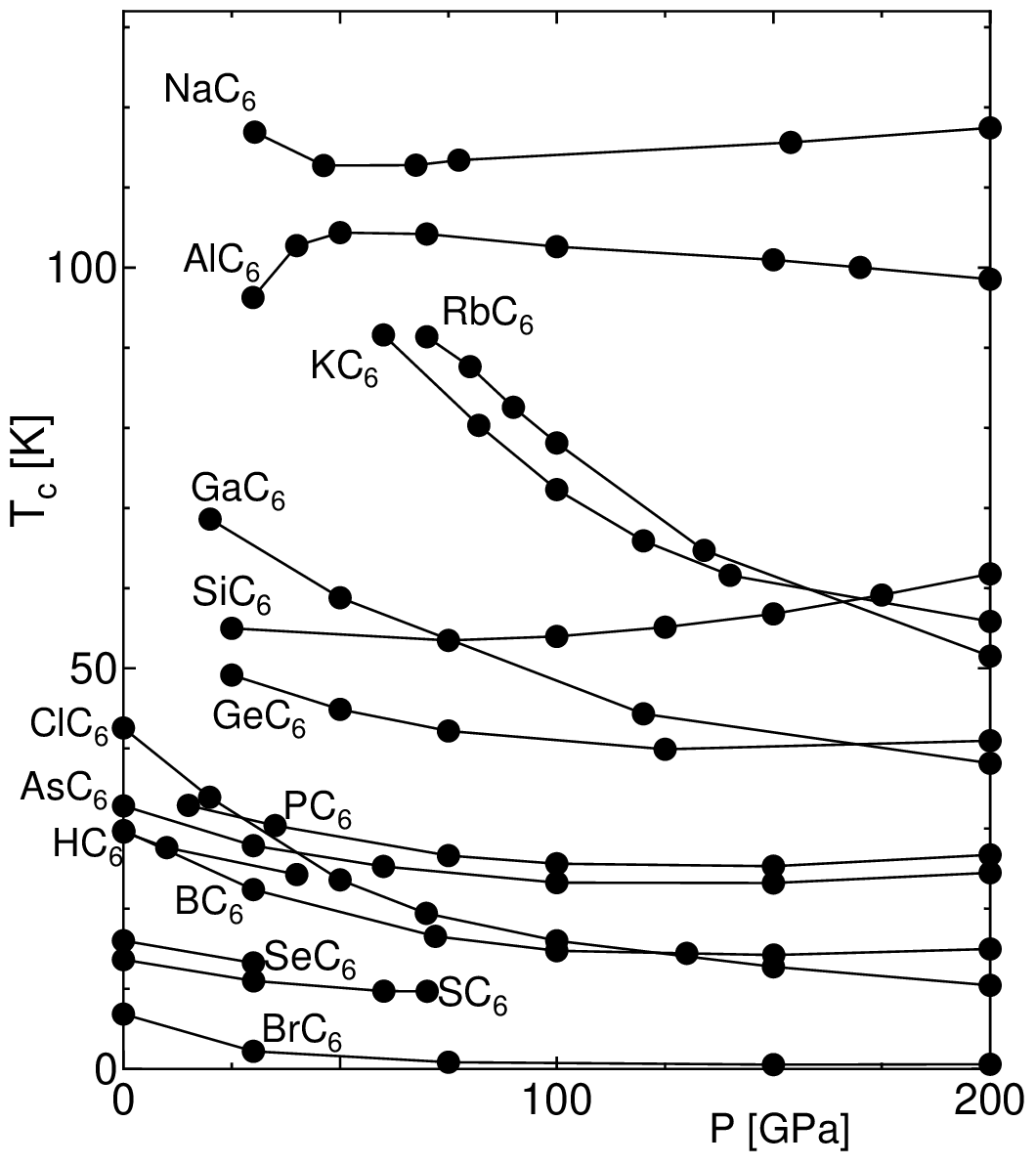}
\end{center}
\caption{  Transition temperature $T_{\rm c}$ of $X$C$_6$ as a function of pressure $P$.
}
\label{P-Tc4}
\end{figure}
%
%
%
Next, we consider compounds of  $X$C$_6$ combining carbon atoms and an $X$ atom.
 As  a typical case,    the results for compounds  NaC$_6$ and  ClC$_6$ are mainly shown.
The optimum pressure for stabilizing  the crystal structure  depends on the kind of the $X$ atom  in the compounds.
For example,   NaC$_6$ is stable  above $\sim$  30  GPa, and  ClC$_6$ is stable  even at   0  GPa, where  $a$ is  4.43  ${\rm \AA}$  in  the former case and 4.62 ${\rm \AA}$ in the latter case.

In Figs. \ref{e-band-all3}(a) and \ref{e-band-all3}(b), we show  the electronic band structures of  NaC$_6$ at 30 GPa   and ClC$_6$ at 0 GPa.
When the compound becomes a metal,  the  atom $X$ can be regarded as a dopant that brings carriers into C$_6$. Here, NaC$_6$ corresponds to  the electron doping case   and ClC$_6$ is the hole doping case.
The overall  behavior of the band of NaC$_6$  is similar to that of C$_6$, as shown in Fig. \ref{C6-band}(a).
It suggests that the band structure is rigid against doping.

On the other hand,  the band of ClC$_6$  near $E_{\rm F}$ is more complex than that of  NaC$_6$.
It seems that carbon  and Cl bands are mixed.
Figures \ref{e-band-all3}(c) and \ref{e-band-all3}(d), show the density of states (DOS)  of electrons for  NaC$_6$ and  ClC$_6$, which are displayed to correspond to  each compound.
They indicate that the contribution of Na to DOS at  $E_{\rm F}$ is almost zero,  whereas  that  of  Cl is large.
For  NaC$_6$,  the role of the Na atom   is only  to provide  carriers to the system of C$_6$.
In the case of   ClC$_6$,  the DOS of the Cl atom seems to occupy about half  of the total DOS  near $E_{\rm F}$. However, the overall behavior of DOS is roughly similar to that of    NaC$_6$.  This suggests that the rigid band structure may also be valid  for ClC$_6$.

In Fig. \ref{a2F}, we show the electron-phonon coupling $\lambda (\omega)$ as a function of $\omega$ and phonon DOS  for  NaC$_6$ at 30 GPa and  C$_6$ with -0.9$|e|$ at 0 GPa.
Here,    $\lambda (\omega)=2\int_0^\omega \alpha^2 F(\omega')/\omega' {\rm d}\omega'$,
  where $\alpha^2 F(\omega)$  is the electron-phonon spectral function.
 Figure \ref{a2F}(a) shows that the phonon spectrum is clearly divided into two parts: components of   Na  and C$_6$ atoms.
Furthermore, the contribution of phonons to $\lambda$  is also   divided into two parts, and the Na atom component  is about half of the C$_6$ atoms component.
By comparing  Figs. \ref{a2F}(a) and \ref{a2F}(b),  we find that the phonon DOS   are similar to each other  except at low frequencies.
This suggests that   the presence of the Na atom does not significantly affect  the phonon oscillations of the C$_6$ structure.
We obtained  similar results for  other compounds.

%
In Fig. \ref{phase-dia-B}, we show the phase diagram for  $X$C$_6$, which   corresponds to Fig. \ref{phase-dia-A}.  
The result is represented by a double circle with an atomic symbol.  Here,  the superscript of each  symbol indicates the pressure in the unit of  GPa, and the additional superscript `U`   means  its mechanical and/or dynamical instability.
  The horizontal axis is the value  of  effective charge introduced into the part of C$_6$ and the vertical axis  is lattice constant $a$.
The effective charge is calculated using the  partial DOS of  C$_6$,  as shown in Fig. \ref{e-band-all3}(c) or  \ref{e-band-all3}(d).
For example, by integrating  the  DOS   from the bottom of the band  to the point of  $E_{\rm F}$, we  obtain  the absolute value  of carriers introduced into  the C$_6$ part,   which is regarded  as negative charge.

In this figure, almost all $X$C$_6$ systems  in  the upper left area  painted light blue are unstable\cite{AsC6}.
Similarly, systems in  the lower right area  painted   light red are also unstable.
When pressure increases, the  lattice constant $a$ decreases and the stability of  heavily doped $X$C$_6$  near the blue area  increases.   On the other hand,  most systems  near the red area  become  unstable with increasing $P$.
Generally, this phase diagram is similar to  that for  C$_6$ systems  as shown in Fig. \ref{phase-dia-A}.

As shown in Figs. \ref{C6super},   we can expect that the necessary conditions for a higher $T_{\rm c}$ are heavy doping and large $a$.
 On the basis of such expectations,  we show  $T_{\rm c}$ as a function of pressure $P$ in Fig. \ref{P-Tc4}.
It shows that some compounds that are located in   the upper left or the  upper right  areas  in Fig. \ref{phase-dia-B} indicate relatively high-$T_{\rm c}$.
In particular,  $T_{\rm c}$ of NaC$_6$ and AlC$_6$ is over 100 K at a relatively low pressure, and  $T_{\rm c}$ of  ClC$_6$ is  $ \sim 40$K   at 0 GPa. 
This result is  almost consistent with a result of the previous works.\cite{Lu-2016,Khan2022}

%
%
In summary,  we  investigated the mechanical and dynamical instabilities and  superconductivity  of   $X$C$_6$  with a sodalite structure which is similar to  a hydrogen compound  showing HTS.   
The electronic states for the C$_6$ structure  with the fictitious charge and  many compounds $X$C$_6$  were systematically examined  by  the first-principles calculations. 
We  classified  the stability  and the superconductivity of  C$_6$ alone  and  $X$C$_6$ on the plane of   the effective doping charge  versus the lattice constant under   external pressure.  
Although the hydrogen compounds  with  HTS require an extremely high pressure, the carbon compounds are  stable at relatively low pressures,  some even at $P=0$.
Electron-doped-like compounds such as  NaC$_6$ or AlC$_6$  show  $T_{\rm c}$ of up to  $\sim$100 K  at $P \simj 30$  GPa, and the results suggest that the sodalite structure of carbon may be a key to generating phonon-mediated HTS.

\begin{acknowledgment}
This work was supported by JSPS KAKENHI Grant Numbers JP15K05168 and JP19K03716.
\end{acknowledgment}




\begin{thebibliography}{99}
\bibitem{Tc-too-high}
The transition temperature $T_{\rm c}$ is  calculated using  the Allen and Dynes formulation\cite{Allen} for $T_{\rm c}$, where the screened Coulomb potential parameter  $\mu^*$  is  set to $0.1$ as a typical value.
\bibitem{AsC6}
As shown in Fig. \ref{phase-dia-B}, AsC$_6$  is stable at $P=0$ even in the blue area   and it seems to be an exception.
Furthermore, our analysis  based on Fig.\ref{phase-dia-A} does not seem to be appropriate for   $X'$C$_6$,  where   $X'$ is Li,  Mg, or another  alkaline earth metal(not shown).
Probably, these elements  have a significant effect on C$_6$ in addition to the effect of doping.
\end{thebibliography}
\end{document}